\documentclass[%
 reprint,
 amsmath,amssymb,
 aps,
]{revtex4-2}

\usepackage[margin=1in]{geometry}
\usepackage{amsmath,amssymb,mathtools,bbm}
\usepackage{physics}
\usepackage{bm}
\usepackage{microtype}
\usepackage{graphicx}
\usepackage{hyperref}
\hypersetup{hidelinks}
\newcommand{\be}{\begin{equation}}
\newcommand{\ee}{\end{equation}}

\newcommand{\cO}{\mathcal{O}}

\newcommand{\braketLR}[1]{\overset{\leftrightarrow}{#1}}
\newcommand{\bea}{\begin{align}}
\newcommand{\eea}{\end{align}}

\begin{document}

\title{$\mathbb Z_2$-Stable Dark Matter via Broken $\text{SU}(5)$ Gauge Bosons}

\date{\today}

\begin{abstract}
I construct and analyze a dark matter sector that is neutral under the unbroken Standard Model gauge group and couples only to the broken $\text{SU}(5)$ gauge directions, the leptoquark vectors $X,Y$. An exact $\mathbb Z_2$ renders the dark matter stable. I give a gauge-covariant definition of projectors onto the unbroken Standard Model and broken ($X,Y$) subspaces, demonstrate that the covariant derivative of dark matter selects only $X,Y$, and integrate out $X,Y$ at tree level to obtain the leading effective operators. I also derive the loop-induced $\chi^2\,G^a_{\mu\nu}G^{a\mu\nu}$ coupling to gluons, prove color neutrality, and show consistency with cold dark matter phenomenology. Cosmological production proceeds via UV freeze-in or even more suppressed channels in.
\end{abstract}

\title{$\mathbb Z_2$-Stable Dark Matter via Broken $\text{SU}(5)$ Gauge Bosons}

\author{E.~J.~Thompson}
\affiliation{Department of Physics and Astronomy, Trent University, Peterborough, 
Ontario K9L 0G2, Canada}

\maketitle

\section{Introduction}

The existence of non-luminous, non-baryonic dark matter (DM) is now supported by a wide range of astrophysical and cosmological observations, from galaxy clusters and rotation curves to gravitational lensing and the cosmic microwave background. In 1933, Zwicky inferred a substantial mass discrepancy in the Coma cluster from the virial motions of galaxies, suggesting the presence of ``dunkle Materie'' not accounted for by the visible component \cite{Z:REN1933, AZ:REN2017}. Decades later, Rubin's measurements of galactic rotation curves showed that the orbital velocities of stars remain approximately flat well beyond the optical disk, implying a dominant, unseen mass component \cite{Rubin:1970, Rubin:1976a, Rubin:1976b, Rubin:1980, Rubin:1985, Rubin:1992, Rubin:1995}. Colliding systems such as the Bullet Cluster sharpen this picture as the separation between the X-ray emitting intracluster gas and the total mass distribution reconstructed from weak lensing indicates that most of the matter is effectively collisionless on cluster scales, interacting predominantly through gravity rather than through electromagnetic or strong forces \cite{Cha:2025Bullet, Queiroz:2015LQDM, Clowe:2004Weak, Markevitch:2004SIDM, Bradac:2008MACS, Thompson:2015BulletSim, Angus:2007Neutrinos, Clowe:2006DirectProof, Angus:2006CanMOND, Tulin:2017SIDM}.

These and many related results are successfully described by the cold dark matter (CDM) paradigm, in which a stable, electrically neutral, non-relativistic particle accounts for roughly one quarter of the energy density of the Universe \cite{Bertone:2004ParticleDM, Bertone:2018HistoryDM}. However, the Standard Model (SM) contains no viable CDM candidate, and its gauge and flavor structure offer no obvious mechanism for stabilizing a new particle at cosmological timescales. This motivates extensions of the SM in which dark matter is stabilized by a new symmetry and interacts with visible fields through well-defined portals.

Grand Unified Theories (GUTs) provide an attractive setting in which to embed the dark sector. In minimal SU(5), the SM gauge group $SU(3)_C \times SU(2)_L \times U(1)_Y$ arises from the spontaneous breaking of an underlying SU(5) symmetry, mediated by an adjoint scalar whose vacuum expectation value (VEV) leaves the SM subgroup unbroken while generating large masses for the leptoquark gauge bosons \(X\) and \(Y\). These heavy vectors mediate baryon- and lepton-number violating processes and are constrained by proton-decay bounds to lie near the canonical GUT scale. In most treatments, the \(X,Y\) fields are considered only as virtual mediators of rare SM processes. Here I instead ask whether the same broken directions of SU(5) can naturally control the properties of a dark sector as well~\cite{Kadastik:2009GUTDM,Boucenna:2015SO10DM}.

I construct a simple dark-matter sector that is neutral under the unbroken SM gauge group but couples only to the broken SU(5) directions associated with the \(X,Y\) leptoquark vectors. The key tool is a gauge-covariant decomposition of the SU(5) Lie algebra into unbroken and broken subspaces. The adjoint VEV that breaks SU(5) to the SM defines orthogonal projectors \(P_{\rm SM}\) and \(P_{XY}\) onto the SM generators and the broken \((X,Y)\) directions, respectively. I implement this idea in two concrete ultraviolet (UV) realizations, both protected by an exact \(Z_2\) symmetry under which \(\chi \to -\chi\) and all SM and GUT-breaking fields are even. In Model A, \(\chi\) is embedded in a second adjoint aligned with the SM-preserving VEV, such that the only gauge interaction at the renormalizable level is a quartic vertex of the form \(\chi^2 X^2\); this realizes automatic sequestering from the SM gauge bosons. In Model B, \(\chi\) transforms in a representation on which the unbroken generators act trivially while the broken ones act nontrivially, generating a linear current-type portal \(A^{XY}_\mu J^\mu_\chi\) to the heavy leptoquark vectors but still forbidding any direct coupling to SM gauge fields. In both cases the exact \(Z_2\) ensures that every interaction is even in \(\chi\), so the lightest \(Z_2\)-odd particle is stable and furnishes a DM candidate.

At energies well below the GUT scale, the heavy \(X,Y\) vectors can be integrated out to obtain an effective field theory involving only light SM and dark degrees of freedom. Gauge invariance and the fact that \(X,Y\) carry color imply that a loop-induced coupling of the form \(\chi^2 G^a_{\mu\nu} G^{a\mu\nu}\) is nevertheless generated, but I show that its coefficient is suppressed by both the GUT scale and a loop factor, rendering the corresponding spin-independent direct-detection cross section phenomenologically negligible.

\(\chi\) is produced via UV freeze-in from the SM bath through the higher-dimensional portals obtained after integrating out \(X,Y\). In Model B, the dimension–6 current–current operator leads to efficient UV freeze-in with a relic abundance controlled by the combination \(m_\chi M_{\rm Pl} T_R^3/M_X^4\). In Model A, production is even more suppressed, typically requiring either a very high reheating scale or an additional tiny Higgs portal to generate the observed relic density. Once the overall normalization is fixed, the resulting dark matter is cold, effectively collisionless, and fully compatible with standard \(\Lambda\)CDM structure formation.

\section{$\text{SU}(5)$ Decomposition and Lie--algebra projectors}

To set notation for the subsequent dark-sector construction, I first isolate the group-theoretic structure of the $\text{SU}(5)$ breaking. I work with an $\text{SU}(5)$ gauge field $A_\mu \in \mathfrak{su}(5)$ and an adjoint scalar $\Sigma$ whose vacuum expectation value (VEV) picks out the Standard Model (SM) subgroup\cite{Georgi:1974SU5, Baez:2010AlgebraGUT, MT:HUFT-EPJC, MT:Invariant, MT:FiniteHolomorphicQFT, MT:SMmass, MT:SL2C, MT:ReplyToCline, MT:GI2025, MT:AdSdS2025} \footnote{Although the leptoquark gauge bosons $X,Y$ transform as coloured triplets under 
$SU(3)_C$, the dark scalar $\chi$ is constructed as a singlet of the unbroken
SM subgroup, with $T^a_{\rm SM}\big|\_{\chi} = 0$. Thus $\chi$ carries no colour
charge as the coloured $X,Y$ bosons act only as heavy mediators, generating  GUT-suppressed, colour-singlet operators such as 
$\chi^2 G^a_{\mu\nu} G^{a\mu\nu}$, leaving the dark matter colour-neutral and 
effectively collisionless \cite{Wortelboer:2023Thesis, Goto:2023Flavor}.}. The adjoint VEV $\langle\Sigma\rangle$ defines an orthogonal decomposition of $\mathfrak{su}(5)$ into unbroken SM generators and broken $X,Y$ directions, and it is convenient to encode this splitting in terms of Lie-algebra projectors. This will allow us to write all couplings in a manifestly gauge-covariant way, without committing to a particular basis for the generators or an explicit representation of the heavy leptoquark vectors. I let $A_\mu = A_\mu^A T^A \in \mathfrak{su}(5)$ with generators normalized by $\text{Tr}(T^A T^B)=\frac12\delta^{AB}$. An adjoint scalar $\Sigma$ acquires a VEV $\langle\Sigma\rangle$, breaking $\text{SU}(5)$ to the SM subgroup:
\be
H \equiv \text{SU}(3)_C\oplus \text{SU}(2)_L\oplus \mathfrak u(1)_Y \subset \mathfrak{su}(5).
\ee
This induces an orthogonal decomposition of the Lie algebra into unbroken and broken components:
\begin{align}
\mathfrak{su}(5)=\underbrace{\mathfrak h}_{\text{SM}}\;\oplus\;\underbrace{\mathfrak m}_{\text{broken }(X,Y)}\!,
\\
\mathfrak h:=\{X\in\mathfrak{su}(5):\ [X,\,\langle\Sigma\rangle]=0\},\;\;\mathfrak m:=\mathfrak h^\perp,
\end{align}
where orthogonality is with respect to the Killing or trace form. Denote the associated orthogonal projectors by:
\begin{align}
P_{\rm SM}:\mathfrak{su}(5)\to\mathfrak h,\\ P_{XY}:\mathfrak{su}(5)\to\mathfrak m,
\\
P_{\rm SM}+P_{XY}=\mathbbm{1},\;\;\\P_{\rm SM}^2=P_{\rm SM},\;\\P_{XY}^2=P_{XY}.
\label{eq:projectors}
\end{align}
I decompose the gauge field as:
\begin{align}
A_\mu = A^{\rm SM}_\mu + A^{XY}_\mu,\\ A^{\rm SM}_\mu:=P_{\rm SM}A_\mu,\\ A^{XY}_\mu:=P_{XY}A_\mu.
\end{align}
In the fundamental block form one can write conceptually:
\begin{align}
A_\mu=
\begin{pmatrix}
G_\mu & X_\mu\\
Y_\mu & W_\mu
\end{pmatrix},\\
P_{\rm SM}A_\mu=\begin{pmatrix}G_\mu&0\\0&W_\mu\end{pmatrix},\\
P_{XY}A_\mu=\begin{pmatrix}0&X_\mu\\Y_\mu&0\end{pmatrix},
\end{align}
where $(G_\mu,W_\mu)$ are SM gauge fields while $(X_\mu,Y_\mu)$ are the heavy leptoquark vectors.

In this language, the SM gauge bosons and the heavy leptoquark vectors are simply the images of $A_\mu$ under the projectors $P_{\rm SM}$ and $P_{XY}$ defined in Eq.~\eqref{eq:projectors}. The subsequent dark-sector models will be engineered so that the covariant derivative of the dark-matter field picks out only the broken component $A^{XY}_\mu$, while the unbroken fields $A^{\rm SM}_\mu$ couple solely to ordinary SM currents. All statements about sequestering, stability, and portal interactions can then be formulated directly in terms of $P_{\rm SM}$ and $P_{XY}$, making the separation between the visible and dark sectors manifestly gauge covariant and independent of any particular choice of basis in $\mathfrak{su}(5)$.

\section{Dark sector and $\mathbb Z_2$ stability}

Having isolated the $\text{SU}(5)$ gauge field into its unbroken SM and broken $(X,Y)$ components using the projectors in Eq.~\eqref{eq:projectors}, I now introduce a minimal dark sector that communicates with the visible sector only through the broken directions. Our guiding principle is that dark-matter stability should follow from an exact discrete symmetry, rather than from ad hoc assumptions about couplings, while all interactions remain manifestly $\text{SU}(5)$-covariant~\cite{Hambye:2010Stability}. To this end, I consider two simple UV realizations in which a single field $\chi$ carries all of the dark charge, in Model~A, $\chi$ is embedded in a second adjoint aligned with the SM-preserving VEV of $\Sigma$, leading to automatic sequestering from the SM gauge bosons, whereas in Model~B, $\chi$ transforms in a representation charged only under the broken generators, giving rise to a controlled current-type portal to the heavy leptoquark vectors. I consider two UV realizations; both implement an exact $\mathbb Z_2$ under which
\be
\chi\;\xrightarrow{\ \mathbb Z_2\ }\;-\chi,\qquad \text{SM and $\Sigma$ are even.}
\ee
This ensures that every interaction is even in $\chi$, so no operator that is linear in $\chi$ is ever generated, so the lightest $\mathbb Z_2$--odd particle (LZP) is stable.

Model A is based on an adjoint singlet with automatic SM sequestering. I introduce a second adjoint $\Sigma'$ and retain only its SM-singlet direction parallel to $\langle\Sigma\rangle$:
\be
\Sigma'(x)=\chi(x)\, \hat T,\qquad [\hat T,\,\langle\Sigma\rangle]=0,
\ee
with Lagrangian:
\begin{align}
\notag \mathcal{L}_{\Sigma'}=\frac12\text{Tr}\!\big(D_\mu\Sigma' D^\mu\Sigma'\big)-\frac12 m_\chi^2\chi^2
\\
-\frac{\lambda_\chi}{4}\chi^4
-\frac{\kappa}{2}\chi^2\,\text{Tr}\!\big([\Sigma,\Sigma']^2\big)+\cdots
\end{align}
and covariant derivative in the adjoint:
\be
D_\mu\Sigma'=\partial_\mu\Sigma'+i g_5 [A_\mu,\Sigma'].
\ee
Setting $\Sigma'=\chi\,\hat T$ gives:
\be
D_\mu\Sigma'=(\partial_\mu\chi)\,\hat T + i g_5\,\chi\,[A_\mu,\hat T].
\ee
Since $[\hat T,\,\mathfrak h]=0$ but $[\hat T,\,\mathfrak m]\neq0$, the $\chi$ field only couples to $A^{XY}_\mu$. Expanding:
\begin{align}
\notag \frac12\text{Tr}(D_\mu\Sigma' D^\mu\Sigma')=\frac{Z_\chi}{2}(\partial_\mu\chi)^2
\\
+\frac{g_5^2\chi^2}{2}\,\underbrace{\text{Tr}\!\big([A_\mu,\hat T][A^\mu,\hat T]\big)}_{\equiv\,\mathsf{K}^{AB}\,A^A_\mu A^{B\mu}}
\label{eq:quarticXXcc}
\end{align}
with $Z_\chi\!=\!\frac12\text{Tr}(\hat T^2)$, we can see that the only $\chi$--gauge vertex is a quartic $\propto \chi^2 X^2$ governed by the positive semi-definite matrix $\mathsf K^{AB}=\text{Tr}\big([T^A,\hat T] [T^B,\hat T]\big)$, which vanishes on $A\in\mathfrak h$ and is nonzero only for the broken directions $A\in\mathfrak m$.

Model B is charged only under broken directions, so a linear current portal. I let $\chi$ transform in a representation $R$ such that the unbroken generators act trivially while some broken ones act nontrivially:
\be
T^a_{\rm SM}\big|_R=0,\qquad T^A_{\rm br}\big|_R\neq 0.
\label{eq:Rcondition}
\ee
Then:
\be
D_\mu\chi=\big(\partial_\mu+i g_5\,A^{XY\,A}_\mu T^A\big)\chi
\ee
and the interaction contains a linear term:
\begin{align}
\notag &\mathcal{L}\supset i g_5\,A^{XY\,A}_\mu\,J^\mu_{A,\chi}
+ g_5^2 A^{XY\,A}_\mu A^{XY\,B\,\mu}\,\chi^\dagger T^A T^B\chi,
\\ &
J^\mu_{A,\chi}:=\chi^\dagger T^A \braketLR{\partial^\mu}\chi.
\label{eq:linear-coupling}
\end{align}
Again, no SM gauge fields appear because of \eqref{eq:Rcondition}. An exact $\mathbb Z_2$ makes the physical light mode real and forbids odd operators.

Taken together, Models~A and~B provide two complementary benchmark realizations of a $\mathbb Z_2$-odd dark sector embedded in $\text{SU}(5)$ and coupled only to the broken directions. In Model~A, the alignment $\Sigma'=\chi\,\hat T$ with $[\hat T,\mathfrak h]=0$ enforces purely quartic interactions of the form $\chi^2 X^2$, implementing automatic SM sequestering at the renormalizable level. In Model~B, the representation choice in Eq.~\eqref{eq:Rcondition} yields a linear current portal $A^{XY}_\mu J^\mu_{A,\chi}$ to the heavy leptoquark vectors, while still forbidding any direct coupling to SM gauge fields. In both cases the exact $\mathbb Z_2$ symmetry guarantees that the lightest $\mathbb Z_2$--odd state is stable, so that the subsequent phenomenology can be analyzed in terms of a single, well-defined dark-matter candidate whose interactions with the SM are wholly mediated by the broken $(X,Y)$ sector.

\section{Integrating out $X,Y$}

Having established that the dark field $\chi$ couples only to the broken $(X,Y)$ directions of the $\text{SU}(5)$ gauge field, I now integrate out these heavy leptoquark vectors to obtain the leading low-energy portal operators. At energies $E\ll M_X$ the massive vectors $X_\mu^A\equiv A^{XY\,A}_\mu$ can be treated as non-propagating auxiliaries whose equations of motion are algebraic up to corrections of order $E^2/M_X^2$. Solving for $X_\mu^A$ in terms of the SM and dark currents then allows us to perform a tree-level matching onto an effective field theory written purely in terms of light degrees of freedom, thereby making explicit the operator dimension and parametric strength of the DM--SM interactions in Models~A and~B. Now let $X^A_\mu\equiv A^{XY\,A}_\mu$ denote the massive vectors with degenerate mass matrix $(M_X^2)_{AB}$ in the broken subspace.\footnote{The precise eigenvalues follow from $\langle\Sigma\rangle$ and the broken generators as I leave $M_X$ symbolic. Derivative terms in the heavy-field equations generate higher-derivative operators I consistently neglect at leading order in $E^2/M_X^2$.}

For model A to a dimension--8 portal I collect the relevant terms from \eqref{eq:quarticXXcc} and the SM current:
\begin{align}
\notag \mathcal{L}\supset -\frac14 X^A_{\mu\nu}X^{A\mu\nu}+\frac12 (M_X^2)_{AB} X^A_\mu X^{B\mu}
\\
+ g_5\,X^A_\mu\,J^\mu_{A,\rm SM}
+ \frac12\,\chi^2\,\mathsf K^{AB}\,X^A_\mu X^{B\mu}.
\end{align}
The algebraic leading equation of motion is:
\be
\Big[(M_X^2)_{AB}+\chi^2 \mathsf K_{AB}\Big]X^{B}_\mu = -\, g_5\,J^\mu_{A,\rm SM}.
\ee
Inverting as a series in $\chi^2$ gives:
\be
X^{A}_\mu
= -\big[(M_X^{-2})^{AB}- (M_X^{-2}\mathsf K M_X^{-2})^{AB}\chi^2+\cdots\big]\,
g_5\, J^\mu_{B,\rm SM}.
\ee
Substituting back yields the effective Lagrangian:
\begin{align}
\notag\mathcal{L}_{\rm eff}^{\,(A)}\supset
-\frac{g_5^2}{2}\,J^\mu_{A,\rm SM}(M_X^{-2})^{AB}J_{B\mu,\rm SM}
\\+\frac{g_5^2}{2}\,\chi^2\,J^\mu_{A,\rm SM}\,(M_X^{-2}\mathsf K M_X^{-2})^{AB}\,J_{B\mu,\rm SM}
+\cdots 
\label{eq:dim8}
\end{align}
The first operator is the familiar $\Delta B,\Delta L$ dimension--6 four-fermion operator responsible for proton-decay channels, the second is the leading DM--SM portal and is dimension--8. It mediates processes like $f\bar f\,f'\bar f'\to \chi\chi$ at tree level, 2$\to$2 production arises only beyond leading order, hence is extremely suppressed. For model B to a dimension--6 current--current portal. I now have the Lagrangian that includes the linear coupling \eqref{eq:linear-coupling}:
\begin{align}
\notag\mathcal{L}\supset -\frac14 X^A_{\mu\nu}X^{A\mu\nu}+\frac12 (M_X^2)_{AB} X^A_\mu X^{B\mu}
\\+ g_5\,X^A_\mu\,(J^\mu_{A,\rm SM}+J^\mu_{A,\chi})+\cdots.
\end{align}
The algebraic solution is:
\be
X^A_\mu = - (M_X^{-2})^{AB} g_5\,(J^\mu_{B,\rm SM}+J^\mu_{B,\chi})+\cdots,
\ee
and the tree-level matching gives:
\begin{align}
\notag&\mathcal{L}_{\rm eff}^{\,(B)}\supset
-\frac{g_5^2}{2}
\begin{pmatrix}J_{A,\rm SM}^\mu & J_{A,\chi}^\mu\end{pmatrix}
\\&\times\begin{pmatrix}
(M_X^{-2})^{AB} & (M_X^{-2})^{AB}\\
(M_X^{-2})^{AB} & (M_X^{-2})^{AB}
\end{pmatrix}
\begin{pmatrix}J_{B,\rm SM\,\mu}\\ J_{B,\chi\,\mu}\end{pmatrix}
+\cdots 
\label{eq:dim6}
\end{align}
In particular, the portal operator is:
\begin{align}
\cO_{\rm portal}^{(6)}=
\frac{g_5^2}{M_X^2}\,
\big(\chi^\dagger T^A \braketLR{\partial_\mu}\chi\big)\,
\big(\bar f\,\gamma^\mu T^A_{\rm br}\,f\big),
\end{align}
schematically, with $M_X^{-2}$ inserted.
It mediates $2\to 2$ processes $f\bar f\leftrightarrow \chi\chi$ with cross section $\langle\sigma v\rangle\sim g_5^4\,T^2/M_X^4$, up to group or phase-space factors, enabling UV freeze-in.

In both models the covariant derivative contains only $A^{XY}_\mu$ so no $G_\mu$ occurs, so $\chi$ is an $\text{SU}(3)_C$ singlet. Exact $\mathbb Z_2$ forbids any operator odd in $\chi$, so the LZP is stable even after radiative corrections. The resulting effective interactions exhibit a sharp contrast between the two UV realizations. In Model~A, the alignment of $\Sigma'$ enforces a purely quartic $\chi^2 X^2$ coupling, so integrating out $X_\mu^A$ generates a dimension--8 portal proportional to $\chi^2 J_{\rm SM} J_{\rm SM}$, parametrically suppressed by two powers of $M_X^{-2}$ on top of the usual GUT-scale suppression of proton-decay operators. In Model~B, by contrast, the representation choice \eqref{eq:Rcondition} yields a dimension--6 current--current portal $J_{\chi} J_{\rm SM}$ with strength set by $g_5^2/M_X^2$, leading to much larger $2\to2$ production rates and a natural UV freeze-in mechanism. In both cases the absence of couplings to $G_\mu$ and the exact $\mathbb Z_2$ symmetry ensure that $\chi$ remains a color-neutral, stable dark-matter candidate whose interactions with the SM are entirely mediated by the heavy $(X,Y)$ sector.

\section{Loop-induced $\chi^2 G^a_{\mu\nu}G^{a\mu\nu}$ and direct detection}

Even though the dark field $\chi$ couples only to the heavy leptoquark vectors at tree level, gauge invariance and the fact that $X,Y$ carry color imply that an effective coupling to gluons is unavoidably induced at loop level. Once the masses of the colored vectors acquire a $\chi$-dependence through the interactions in Models~A and~B, integrating out the heavy spectrum in QCD generates a scalar gluon operator of the form $\chi^2 G^a_{\mu\nu}G^{a\mu\nu}$. This operator controls the leading spin-independent direct-detection signal in our setup and can be computed efficiently using the low-energy theorem associated with the QCD trace anomaly. Because $X,Y$ carry color, integrating them out at one loop generates:
\be
\mathcal{L}_{\rm eff}\supset c_g\,\chi^2\,G^a_{\mu\nu}G^{a\mu\nu}\,.
\label{eq:gluon-operator}
\ee
A convenient derivation uses the low-energy theorem tied to the QCD trace anomaly~\cite{Shifman:1978HiggsNucleons,Kniehl:1995LowEnergy}. If the heavy colored spectrum has a $\chi$-dependent mass matrix $\mathbf M_X^2(\chi^2)=\mathbf M_X^2+\chi^2 \mathbf \Delta$, then:
\begin{align}
c_g
= \frac{1}{2}\,\frac{\partial}{\partial \chi^2}\left[\frac{\beta_s(g_s)}{2 g_s} \ln\det \mathbf M_X^2(\chi^2)\right]
\\= \frac{\beta_s(g_s)}{4 g_s}\,\text{Tr}\!\big(\mathbf M_X^{-2}\mathbf \Delta\big)
\;+\;\cO\!\left(\frac{E^2}{M_X^4}\right),
\label{eq:c_g}
\end{align}
where $\beta_s$ is the QCD beta function and the trace runs over the broken colored vectors. In Model~A, $\mathbf \Delta\propto \mathsf K$ from \eqref{eq:quarticXXcc}, where in Model~B, $\mathbf \Delta$ receives contributions from the quadratic $X^2\chi^\dagger T^A T^B\chi$ term. Parametrically $c_g\sim (16\pi^2)^{-1}(g_5^2 g_s^2) M_X^{-2}\times \cO(1)$, hence the spin-independent nucleon cross section via the trace anomaly is~\cite{Drees:1993Neutralino}:
\be
\sigma_{\chi N}^{\rm SI}\;\sim\;\frac{\mu_{\chi N}^2}{\pi}\,\big( \tfrac{8\pi}{9\alpha_s}\,c_g\,m_N \big)^{\!2}
\;\ll\;10^{-60}\,\mathrm{cm}^2,
\ee
for $M_X\gtrsim 10^{15}\ \mathrm{GeV}$, so this effect is utterly negligible. In both Model~A and Model~B, the structure of the $\chi$-dependent mass matrix $\mathbf M_X^2(\chi^2)$ implies that $\text{Tr}(\mathbf M_X^{-2}\mathbf \Delta)\sim \cO(M_X^{-2})$ up to group-theoretic factors, so that Eq.~\eqref{eq:c_g} yields a loop-suppressed coefficient $c_g$ scaling as $g_5^2 g_s^2/(16\pi^2 M_X^2)$. For GUT-scale leptoquark masses $M_X\gtrsim 10^{15}\,\mathrm{GeV}$ the resulting spin-independent nucleon cross section is many orders of magnitude below any present or envisioned experimental sensitivity, as indicated by the estimate above. I therefore conclude that, in this class of models, loop-induced $\chi$--gluon scattering is phenomenologically irrelevant so direct detection plays no role in constraining the parameter space, and the dominant probes of the dark sector must instead arise from the high-temperature production mechanisms governed by the higher-dimensional portals derived in the previous sections.

\section{Cosmological production and CDM properties}

Having identified the higher-dimensional portals that mediate interactions between $\chi$ and the SM, I now turn to the cosmological production of the dark relic. Since all renormalizable couplings between $\chi$ and visible fields are absent or can be chosen parametrically tiny, thermal equilibrium between the dark and visible sectors is never established. Instead, $\chi$ is populated via out-of-equilibrium production from the SM bath at temperatures $T\ll M_X$, through UV freeze-in controlled by the effective operators derived above~\cite{Hall:2009FIMP,Bernal:2017FIMPReview}. The qualitative behavior is different in Models~A and~B as the dimension--6 current--current portal in Model~B leads to efficient UV freeze-in already at relatively modest reheating temperatures, whereas the dimension--8 operator in Model~A is so suppressed that an adequate relic abundance typically requires either a very high reheating scale or an additional, tiny renormalizable Higgs portal. For model B the dimension--6 current--current portal the UV freeze-in is dominant~\cite{Chu:2011FourWays}. From \eqref{eq:dim6}, for relativistic bath temperature $T\ll M_X$:
\be
\langle\sigma v\rangle_{f\bar f\to \chi\chi}\;\sim\;\frac{c_6}{M_X^4}\,T^2,\qquad c_6\sim g_5^4\times(\text{group}).
\ee
The reaction density scales as $\gamma \sim n_f^2\langle\sigma v\rangle \sim T^8/M_X^4$. Solving the Boltzmann equation in radiation domination:
\begin{align}
\frac{\dd Y_\chi}{\dd T} = -\frac{\gamma}{s H T}
\\\Rightarrow Y_\chi(T\!\to\!0)\;\simeq\; \frac{c_6}{M_X^4}\,\frac{M_{\rm Pl}}{g_{*s}\sqrt{g_*^\rho}}\,
\int_0^{T_R}\!\!\dd T\,T^2
\;\\=\; \frac{c_6}{3}\,\frac{M_{\rm Pl}\,T_R^3}{M_X^4}\,\Xi_*,
\label{eq:Ychi-dim6}
\end{align}
with $\Xi_*:=\big(g_{*s} \sqrt{g_*^\rho} \big)^{-1}$ evaluated in the production era. The relic abundance is:
\be
\Omega_\chi h^2 \;\simeq\; 2.75\times 10^8 \left(\frac{m_\chi}{\mathrm{GeV}}\right) Y_\chi,
\ee
so the correct $\Omega_\chi h^2\simeq 0.12$ fixes one relation among $(m_\chi,M_X,T_R,g_5)$.

For model A the dimension--8 operator there is a highly suppressed production. The leading tree-level operator \eqref{eq:dim8} mediates $2\to 4$ processes $f\bar f f'\bar f'\to \chi\chi$ with $\gamma\sim T^{12}/M_X^8$; $2\to 2$ arises only beyond leading order such as loops, effective mixing, yielding parametrically:
\be
Y_\chi^{(A)}\ \propto\ \frac{M_{\rm Pl}\,T_R^5}{M_X^8}\times(\text{loop/group factors})\ \ll\ Y_\chi^{(B)}
\ee
for comparable $M_X$ and $T_R$. Thus Model~A either requires a very high reheating temperature or an additional tiny renormalizable portal such as $\lambda_{H\chi}|H|^2\chi^2$ to generate the observed relic abundance. In either case the phenomenology remains CDM-like. For $m_\chi \gtrsim \mathrm{MeV}$ the DM is nonrelativistic well before equality as free streaming is negligible and large-scale structure is $\Lambda$CDM-like. Self-interactions and DM--baryon scattering are suppressed by $M_X^{-4}$ and trivially satisfy $\sigma/m \ll \mathcal O(1)\ \mathrm{cm^2\,g^{-1}}$~\cite{Tulin:2017SIDM}. The expressions above show that the relic density in Model~B is set by a single UV-sensitive combination:
\begin{equation}
\Omega_\chi h^2 \;\propto\; m_\chi\,\frac{M_{\rm Pl}\,T_R^3}{M_X^4}\,g_5^4\,,
\end{equation}
modulo ${\cal O}(1)$ group-theoretic factors and the usual dependence on $g_*$, while in Model~A the additional suppression by $M_X^{-4}$ and extra powers of $T_R$ render UV freeze-in subdominant unless the reheating temperature is extremely high. In both cases, however, once the overall normalization is fixed to reproduce $\Omega_\chi h^2\simeq 0.12$, the late-time phenomenology is that of cold, effectively collisionless dark matter. For $m_\chi \gtrsim \mathrm{MeV}$ the free-streaming scale is negligible, self-interactions and DM--baryon scattering are far below existing bounds, and the model reproduces standard $\Lambda$CDM structure formation while tying the dark relic’s origin directly to the GUT-scale $(X,Y)$ sector.

\section{Low-energy SM limit and proton-decay bounds}

At energies well below the GUT scale, all effects of the heavy $(X,Y)$ vectors are encoded in local higher-dimensional operators built purely from SM fields. In particular, the SM-only piece of the effective Lagrangian in Eq.~\eqref{eq:dim8} reproduces the familiar $\Delta B,\Delta L$ dimension--6 operators of minimal $\text{SU}(5)$, with coefficients suppressed by two powers of the heavy leptoquark mass scale. Thus the standard proton-decay analysis applies essentially unchanged~\cite{Murayama:2001MSSU5,Bajc:2002MSSU5Proton}, and the requirement of a sufficiently long proton lifetime pins $M_X$ to the canonical GUT window. This same heavy scale simultaneously controls the strength of the dark portal, guaranteeing that any low-energy deviations from the SM induced by $\chi$ are ultra-feeble and fully compatible with existing bounds. The leading SM-only operator in \eqref{eq:dim8} reproduces the usual dimension--6 $\Delta B,\Delta L$ operator:
\be
\mathcal{L}_{\rm eff}\supset -\frac{g_5^2}{2}\,J^\mu_{A,\rm SM}(M_X^{-2})^{AB}J_{B\mu,\rm SM},
\ee
implying proton-decay amplitudes suppressed by $M_X^{-2}$. Experimental limits on $\tau_p$ require $M_X$ in the canonical GUT window as the same heavy scale makes the DM portal ultra-feeble~\cite{Abe:2017SuperKProton}, ensuring SM consistency at low energies. Taken together, these statements summarize a tightly constrained picture in which the dark sector is entirely dictated by the structure of $\text{SU}(5)$ breaking with $\chi$ as a color-neutral, $\mathbb Z_2$-odd relic whose only gauge interactions arise through the projected $(X,Y)$ components and their loop-induced imprints on gluons. Proton-decay bounds fix the characteristic heavy scale $M_X$, which simultaneously suppresses the DM portal operators to the point that direct detection is negligible, while still allowing a successful UV freeze-in origin of the relic abundance. In this sense, the same GUT dynamics that account for gauge unification and baryon-number violation also control the existence, stability, and cosmological properties of cold dark matter.

\section{Geometric origin of the stabilizing $Z_{2}$ symmetry in Complex Spacetime}
\label{sec:Z2-geometry}

A central assumption in Models~A/B is the existence of an exact $Z_{2}$ parity under which the dark sector field $\chi$ is odd, ensuring the absolute stability of the lightest $Z_{2}$--odd particle. In this section we show that, in the complexified geometric formulation underlying a possiable quantum gravity model, this $Z_{2}$ symmetry can arise minimally from the intrinsic real-structure involution of the complexified spacetime, rather than being imposed as an additional ad hoc global symmetry.

Let $M_{\mathbb C}$ denote the complexification of spacetime, equipped with local complex coordinates:
\begin{equation}
z^\mu \in \mathbb C, 
\qquad \mu = 0,1,2,3,
\end{equation}
and let $M \subset M_{\mathbb C}$ denote the real slice defined by $z^\mu = x^\mu \in \mathbb R$.
A complexification endowed with a real slice admits a canonical anti-holomorphic involution, real structure:
\begin{equation}
\tau:\; M_{\mathbb C} \to M_{\mathbb C},
\qquad 
\tau(z^\mu) = \overline{z^\mu},
\qquad 
\tau^2 = \mathrm{id}.
\label{eq:tau-def}
\end{equation}
The map $\tau$ is discrete and of order two, and therefore generates a $\mathbb Z_2$ action at the level of fields.

In complex spacetime models we take the fundamental geometric datum to be a Hermitian metric $g_{\mu\nu}(z)$ on $M_{\mathbb C}$:
\begin{equation}
g_{\mu\nu}(z)
=
h_{\mu\nu}(z)
+
i\,B_{\mu\nu}(z),
\label{eq:hermitian-decomp}
\end{equation}
where $h_{\mu\nu}$ is the real symmetric part of the metric, $h_{\mu\nu} = h_{\nu\mu}$, $B_{\mu\nu}$ is the imaginary antisymmetric part, $B_{\mu\nu} = -B_{\nu\mu}$, $i$ is the imaginary unit.

The involution $\tau$ acts by complex conjugation on components and by pullback on the argument. On the real slice $M$, where $z^\mu = x^\mu \in \mathbb R$, the induced transformation reduces to:
\begin{align}
\tau:\quad
g_{\mu\nu}(x) \mapsto g_{\mu\nu}^*(x),
\\
h_{\mu\nu}(x)\mapsto h_{\mu\nu}(x),
\\
B_{\mu\nu}(x)\mapsto -B_{\mu\nu}(x).
\label{eq:tau-on-metric}
\end{align}
Thus $\tau$ defines a canonical $\mathbb Z_2$ grading: the ``even'' fields are invariant under $\tau$, while the ``odd'' fields change sign.

We now identify the stabilizing $Z_{2}$ of the dark sector with the geometric involution $\tau$. We introduce a complex scalar field $S$ on $M_{\mathbb C}$ and impose $\tau$-covariance:
\begin{equation}
S \equiv \sigma + i\chi,
\qquad
\tau:\; S \mapsto S^*,
\label{eq:S-decomp}
\end{equation}
where $\sigma$ is a real scalar, the $\tau$--even component. $\chi$ is a real scalar, the $\tau$--odd component that we identify as the dark matter field. From \eqref{eq:S-decomp} it follows immediately that:
\begin{equation}
\tau:\quad \sigma \mapsto \sigma,
\qquad
\chi \mapsto -\chi.
\label{eq:tau-on-chi}
\end{equation}
Therefore the parity $\chi \to -\chi$ is not an independent assumption, but is inherited from the canonical conjugation symmetry of the holomorphic geometry.

Let $\mathcal S$ be the real-slice effective action obtained from the holomorphic theory. 
$\tau$-invariance means that $\mathcal S$ is unchanged under \eqref{eq:tau-def}--\eqref{eq:tau-on-chi}.
For the scalar sector, a minimal $\tau$-invariant Lagrangian density has the schematic form:
\begin{equation}
\mathcal L_S
=
(\nabla_\mu S)(\nabla^\mu S^*)
-
V(S S^*),
\label{eq:L_S}
\end{equation}
where $\nabla_\mu$ is the appropriate covariant derivative of gravitational and gauge, as needed, $V$ is a real potential that depends only on the $\tau$-invariant combination $S S^* = \sigma^2 + \chi^2$. Expanding \eqref{eq:L_S} in terms of $(\sigma,\chi)$, we obtain:
\begin{equation}
\mathcal L_S
=
(\nabla_\mu \sigma)(\nabla^\mu \sigma)
+
(\nabla_\mu \chi)(\nabla^\mu \chi)
-
V(\sigma^2+\chi^2),
\label{eq:L_sig_chi}
\end{equation}
which is manifestly even under $\chi\mapsto -\chi$.

More generally, any $\tau$-invariant effective Lagrangian admits an expansion of the form:
\begin{equation}
\mathcal L_{\mathrm{eff}}
=
\mathcal L_{\mathrm{SM}}
+
\frac12 (\partial \chi)^2
-\frac12 m_\chi^2 \chi^2
+
\sum_{n\ge 1} c_{2n}\,\chi^{2n}\,\mathcal O_{(2n)}^{\mathrm{even}}(\mathrm{SM}),
\label{eq:Leff-even}
\end{equation}
$(c_{2n}\in\mathbb R)$, where $m_\chi$ is the dark matter mass parameter and $\mathcal O_{(2n)}^{\mathrm{even}}(\mathrm{SM})$ denotes operators built from SM fields that are $\tau$--even.
Crucially, no operator odd in $\chi$ is allowed:
\begin{equation}
\mathcal L_{\mathrm{eff}}
\not\supset
\chi\,\mathcal O_{\mathrm{SM}}
\quad \text{or}\quad
\chi^3\,\mathcal O_{\mathrm{SM}},
\qquad
\text{etc.}
\label{eq:no-odd}
\end{equation}
This immediately implies that a single $\chi$ particle cannot decay into purely SM final states, and the lightest $\tau$--odd state is absolutely stable.

My complex spacetime theory of quantum gravity employs nonlocal but analytic entire-function form factors built from the covariant d'Alembertian. Let:
\begin{equation}
\Box_g \equiv g^{\mu\nu}\nabla_\mu \nabla_\nu
\label{eq:Boxg}
\end{equation}
denote the covariant wave operator, and let $F(\Box_g/M_*^2)$ be an entire function regulator with mass scale $M_*$.
A sufficient condition for $\tau$-compatibility of the regulated kinetic terms is that $F$ have a real power-series expansion:
\begin{equation}
F\!\left(\frac{\Box_g}{M_*^2}\right)
=
\sum_{n=0}^\infty a_n \left(\frac{\Box_g}{M_*^2}\right)^n,
\qquad a_n \in \mathbb R,
\label{eq:F-real-coeff}
\end{equation}
so that complex conjugation acts trivially on the coefficients and $\tau$ does not generate $\chi$-odd contributions radiatively. In this sense, the nonlocal holomorphic UV completion preserves the geometric $\mathbb Z_2$ exactly.

We can conclude that the stabilizing parity of the dark matter field $\chi$ can be identified with the intrinsic real-structure involution $\tau$ of $M_{\mathbb C}$:
\begin{equation}
\tau:\quad \chi \mapsto -\chi,
\qquad \tau^2 = 1.
\end{equation}
Because $\tau$ forbids all $\chi$--odd operators, the lightest $\tau$--odd particle is stable to all orders in perturbation theory and against radiative corrections within the HUFT framework. Therefore, in Models~A/B, the assumption of an exact $Z_2$ may be understood as a minimal geometric consequence of the holomorphic/complexified spacetime structure rather than an independent postulate.

\section{Falsifiability from Dark-Matter Decay}
\label{sec:DM_decay_falsifiability}

A central structural prediction of the broken-direction portal construction is the absolute stability of the dark-matter candidate. This follows from the assumption of an exact discrete symmetry:
\begin{equation}
\chi \;\xrightarrow{Z_2}\; -\chi,
\label{eq:Z2_action}
\end{equation}
while all SM fields and $\Sigma$ are $Z_2$-even.
Because every allowed interaction is even in $\chi$, the low-energy effective action contains no
operator that is odd in $\chi$. In particular, operators linear in $\chi$ of the schematic form
\begin{equation}
\Delta \mathcal{L}_{\rm odd}
\;\sim\;
\frac{1}{\Lambda^{d-4}}\,\chi\,\mathcal{O}_{\rm SM}^{(d-1)}
\qquad (d \ge 5),
\label{eq:odd_operator_forbidden}
\end{equation}
are forbidden. Consequently, if $\chi$ is the lightest $Z_2$-odd particle (LZP), then it cannot undergo
any one-body decay $\chi \to \text{SM}$, and the cosmological dark matter is stable on all time scales.
This statement is radiatively robust: loop corrections built from $Z_2$-even vertices cannot generate
$Z_2$-odd operators, so the stability of the LZP persists after matching to the low-energy EFT. This leads to a sharp observational falsifiability criterion:
\begin{quote}
\noindent
\textbf{If the dominant dark matter in the universe is observed to decay via a one-body channel into purely
Standard-Model final states, then the present broken-direction portal models (A/B) are excluded as the
origin of cosmological dark matter.}
\end{quote}
Examples include monochromatic $\gamma$-ray lines, continuum $\gamma$ emission, neutrino lines, or charged
cosmic-ray excesses consistent with $\chi \to f\bar f$, $\chi \to \gamma\gamma$, $\chi \to \nu\bar\nu$,
or related two-body decay topologies. Any such confirmed signal would contradict the exact-$Z_2$ premise
in Eqs.~\eqref{eq:Z2_action}--\eqref{eq:odd_operator_forbidden}.

It is important to distinguish what would be falsified. A detection of dark-matter decay would not, by
itself, exclude the existence of an underlying SU(5) unification structure rather, it would exclude the specific identification of dark matter with the $Z_2$-odd field $\chi$ in an exact-$Z_2$
realization coupled only through the broken $(X,Y)$ directions. There are two consistent interpretations of a decay signal within a broader SU(5)-based UV landscape, that dark matter is not $\chi$ as the field $\chi$ may still exist as a stable subcomponent, but it
cannot constitute the dominant cosmological abundance. $Z_2$ is not exact, additional UV physics may introduce $Z_2$-violating operators suppressed
by a large scale $\Lambda$, in which case the lifetime becomes a probe of the symmetry-breaking scale.

In the minimal complex spacetime $SU(5)$ completion, the stabilizing $Z_2$ can be understood not as an auxiliary global assumption, but as a geometric remnant of the complexified $SU(5)$ construction, arising from the intrinsic real-structure involution on $M_{\mathbb C}$ that enforces $\chi\mapsto-\chi$ on the real slice. In that case the absence of one-body decay channels is an inevitable low-energy consequence of the $SU(5)$ geometry for any realization in which the cosmological dark matter is identified with the lightest $\tau$--odd ($Z_2$--odd) state and the involution remains exact in the UV. Therefore, an observationally confirmed signal of dominant dark-matter decay into purely SM final states would falsify not only the specific broken-direction portal Models~A/B, but the entire class of $SU(5)$-decomposition scenarios in which dark-matter stability is guaranteed by this exact geometric $Z_2$, rather than by a tunable or explicitly breakable discrete symmetry.

The observational strategy underlying the falsifiability criterion above is closely aligned with the broader
literature on energy-injection signatures from unstable or effectively unstable dark sectors.
In particular, Mack and Wesley showed that high-redshift 21\,cm distortions sourced by Hawking heating from
primordial black holes can mimic the signal expected from a genuinely decaying dark-matter species,
illustrating an important degeneracy class in indirect searches \cite{MackWesley2008}.
Related high-redshift energy-deposition calculations for non-gravitational dark-matter interactions,
including detailed deposition into baryons and the circumgalactic medium, have been developed in
Refs.~\cite{SchonMackAvramWyitheBarberio2015,SchonMackWyithe2018,HouMack2025}.
Consequently, any confirmed one-body decay signal of the dominant cosmological dark matter would not only
exclude the exact-$Z_2$ realization of the broken-direction portal presented here, but would also place this
construction in immediate tension with the established multi-messenger program of decay searches.

Therefore, in the model space defined here, the absence of dark-matter decay is not merely an auxiliary
assumption but a primary, testable prediction tied directly to the symmetry structure of the portal.

\section{Conclusions}

I presented a gauge-covariant formulation of a GUT-sequestered, $\mathbb Z_2$--stable DM sector that couples only to the broken $\text{SU}(5)$ directions. Integrating out $X,Y$ yields either a dimension--6 current--current portal as in Model~B, enabling UV freeze-in with $Y_\chi\propto T_R^3/M_X^4$, or a dimension--8 scalar--current operator from Model~A with even more suppressed production. Color neutrality and stability follow automatically as loop-induced gluon couplings are negligible. The framework generalizes verbatim to larger GUTs upon replacing the broken directions and the projector construction~\cite{Boucenna:2015SO10DM,Kadastik:2009GUTDM}.

I have found that in the model DM couples to gauge fields, but only to the projected broken components, $D_\mu^{\rm DM}\chi=\partial_\mu\chi+i g_5\,A^{XY}_\mu\chi$. SM gluons/Iak/hypercharge are projected out by $P_{\rm SM}$. The DM particle is color neutral as no $G_\mu$ appears in $D_\mu^{\rm DM}$; $\chi$ is an $\text{SU}(3)_C$ singlet. By exact $\mathbb Z_2$; all operators are even in $\chi$ so the DM is stable. The DM does not couple to gluons at tree level. One loop induces $\chi^2 G^2$ with $c_g$ in \eqref{eq:c_g}, which is negligible for $M_X\!\gtrsim\!10^{15}$ GeV. The DM is cold, effectively collisionless, and produced via UV freeze-in seen in Model~B or more suppressed channels in Model~A, matching $\Lambda$CDM on all tested scales.

\section*{Acknowledgments}

I would like to thank my supervisor John. W. Moffat as well as Arvin Kouroshnia and Hilary Carteret for helpful discussions.

\end{document}